\documentstyle[12pt]{article}
\newcommand {\beq}{\begin{equation}}
\newcommand {\eeq}{\end{equation}}
\newcommand {\beqa}{\begin{eqnarray}}
\newcommand {\eeqa}{\end{eqnarray}}
\newcommand {\n}{\nonumber \\}

\def\pa{\partial}

\begin{document}
\setlength{\oddsidemargin}{0cm}
\setlength{\baselineskip}{7mm}

\begin{titlepage}
 \renewcommand{\thefootnote}{\fnsymbol{footnote}}
$\mbox{ }$
\begin{flushright}
\begin{tabular}{l}
KEK-TH-1113 \\
Nov 2006
\end{tabular}
\end{flushright}

~~\\
~~\\
~~\\

\vspace*{0cm}
    \begin{Large}
       \vspace{2cm}
       \begin{center}
         {Graviton Propagators in Supergravity\\ and Noncommutative Gauge Theory}      \\
       \end{center}
    \end{Large}

  \vspace{1cm}

\begin{center}
{\large 
Yoshihisa K{\sc itazawa}$^{1),2)}$\footnote{kitazawa@post.kek.jp} and
Satoshi N{\sc agaoka}$^{1)}$\footnote{nagaoka@post.kek.jp} 
}

$^{1)}${\it High Energy Accelerator Research Organization (KEK)}\\
{\it Tsukuba, Ibaraki 305-0801, Japan} \\

$^{2)}$ {\it Department of Particle and Nuclear Physics,}\\
                {\it The Graduate University for Advanced Studies,}\\
{\it Tsukuba, Ibaraki 305-0801, Japan}\\
\end{center}

\vfill

\begin{abstract}
\noindent 
\end{abstract}
We investigate the graviton propagator in the type IIB supergravity background
which is dual to 4 dimensional noncommutative gauge theory.
We assume that the boundary is located not at the infinity but 
at the noncommutative scale where the string frame metric exhibits 
the maximum.
We argue that the Neumann boundary condition is the appropriate
boundary condition to be adopted at the boundary.
We find that the graviton propagator behaves just as that of the
4 dimensional massless graviton. 
On the other hand, the non-analytic behaviors of the other Kaluza-Klein modes 
are not significantly affected by the Neumann boundary condition.

\vfill
\end{titlepage}
\vfil\eject

\section{Introduction}
\setcounter{equation}{0}

One of the important questions in string theory is to identify 
the fundamental building blocks of spacetime.  
Although the traditional geometric picture is suited for the closed string
sector, closed/open string duality suggests that a microscopic
description of spacetime may be constructed in terms of branes.
It is certainly conceivable that a macroscopic spacetime emerges
out of a certain brane system. We hope that such a question can be addressed
in matrix model formulations of string theory\cite{IKKT,BFSS}.

As a concrete example, we investigate the 4 dimensional noncommutative spacetime 
which can be regarded as a classical solution of IIB matrix model.
By expanding IIB matrix model around such a noncommutative background,
4 dimensional noncommutative gauge theory with maximum SUSY is obtained
\cite{CDS,AIIKKT,Li}.

The gauge invariant observables in noncommutative gauge theory are 
the Wilson lines\cite{IIKK,Gross,DhKhe}.  
We are most interested in the BPS operators since they couple
to the supergravity degrees of freedom\cite{vertex,ITU}. 
In ordinary gauge theory, the correlation functions of these operators in the real space
decay in accordance with their canonical dimensions.
The characteristic feature of noncommutative gauge theory is the
UV-IR mixing effect which takes place in the non-planar contributions
\cite{MRS}.
In noncommutative gauge theory, the correlators consist of the
planar and non-planar contributions  even in the large $N$ limit. 
It has been found that such an effect takes place
in the two point correlators\cite{KN,KTT}. 

It is natural to expect that these features of the correlation functions 
can be described holographically in terms of supergravity\cite{HI,MR}. 
In string theory, the short distance regime in the
open string sector is mapped to the long distance regime in the closed string sector.
That is why the local operator in gauge theory is assumed to be located at the
boundary of $AdS_5$ where the metric diverges in $AdS$/CFT correspondence
\cite{Maldacena,GKP,Witten}.
In the case of noncommutative gauge theory, the noncommutative scale 
sets the minimum length scale. In fact the string frame metric of the supergravity
background exhibits the maximum at the noncommutativity scale. 
It is consistent with the fact it is not possible to construct more localized states 
than the noncommutative scale.
From these considerations, there has been a proposal to locate the Wilson line operators 
at the maximum of the string frame metric\cite{IIKK2, DhKWL}.  

In this paper, we further investigate this proposal.
We argue that it is natural to adopt the Neumann boundary condition for
the propagators in this proposal. 
With such a prescription, we find that the graviton propagator exhibits the 
$1/p^2$ type pole signaling the existence of 4 dimensional massless gravitons just like
in the Randall-Sundrum model\cite{RS}.
After the introduction,
we briefly recall the Wilson loop correlators in noncommutative gauge theory
in section 2.
In section 3, we investigate the propagators of graviton and Kaluza-Klein modes
in our prescription.
We conclude in section 4 with discussions.
 In Appendix, the propagators of the Kaluza-Klein modes are examined in detail.

\section{Wilson line correlators in noncommutative gauge theory}
\setcounter{equation}{0}

In this section, we briefly recall the correlators of the Wilson lines
in noncommutative gauge theory. 
We may compactify the noncommutative spacetime $R^4$ into the
4 dimensional fuzzy homogeneous spaces $G/H$ in order to regularize the correlators. 
Noncommutative gauge theories on the compact homogeneous spaces $G/H$ 
are derived from matrix models \cite{IKTW,Mathom}. 
They are described by $N \times N$
Hermitian matrices $A_\mu$ where the matrix size $N$ is finite.
Quantum corrections in gauge theories on these spaces
are investigated in \cite{fuzS2,fuzS2S2,KTT1,ABNN1,ABNN2}.
Supersymmetry plays a crucial role to consider the open/closed string duality
since the quantum corrections in the both theories are suppressed 
due to supersymmetry.
$G/H$ spaces restore supersymmetry locally
since these spaces approach the noncommutative flat space in the 
large $N$ limit.
Thus, we focus on compact fuzzy $G/H$ spaces in this section 
and their supergravity 
dual descriptions in the next section.
We analyze the supergravity background 
which is proposed as a dual description of noncommutative Yang-Mills
\footnote{
Supergravity dual descriptions of fuzzy $G/H$ coset spaces are
investigated in \cite{TV}. One can check that a gravity background which is dual
to the noncommutative flat space is obtained from such descriptions in the
large $N$ limit.}.

The important observables are the correlators of the open Wilson lines
in noncommutative gauge theory.
On $G/H$, we construct a generalization of a simple straight Wilson line by the polynomial
of the matrices $A_\mu$ as $tr {\cal Y}_k (A)$ where ${\cal Y}$ denotes 
a spherical harmonics on $G/H$. $k$ denotes the quantum numbers of  a spherical harmonics
which corresponds to a momentum of the  Wilson line. 
The compact fuzzy $G/H$ spaces provide gauge invariant regularizations
for the noncommutative gauge theories.
The bosonic part of the graviton vertex operator is written as 
\begin{eqnarray}
Str {\cal Y}_k [A_\rho,A_\mu][A_\rho,A_\nu] h_{\mu \nu} \ ,
\end{eqnarray}
which shows us the coupling between the open Wilson lines and 
graviton modes $h_{\mu \nu}$.
$\mu$ and $\nu$ run over the directions which are tangent to the space $G$.
For example, on $CP(2)=SU(3)/U(2)$, $\mu, \nu=0,1,\cdots ,7$ 
since $SU(3)$ has 8 generators. On 
$S^2\times S^2 =SU(2)/U(1) \times SU(2)/U(1)$, 
$\mu, \nu =0,1,\cdots ,5$ since each $ SU(2)$ has
3 generators. 

The symmetric trace of the operators on a compact space is defined as
\begin{eqnarray}
Str A^k {\cal O}_1 {\cal O}_2 \equiv \frac{1}{k}
\sum_{k_1=0}^k A^{k_1} {\cal O}_1 A^{k-k_1} {\cal O}_2 \ .
\end{eqnarray}
The operators with this ordering
naturally appear in the construction of the BPS operators\cite{vertex}.
The two point functions of the graviton type vertex operators 
are investigated in \cite{KN}.
The leading nonanalytic behavior, which comes from 
the ultraviolet contribution, is found as
\begin{eqnarray}
\langle Str {\cal Y}_k [A,A][A,A] Str {\cal Y}^\dagger_k 
[A^\dagger,A^\dagger][A^\dagger,A^\dagger]
\rangle \sim \frac{1}{k^2} \ , \label{gravuv}
\end{eqnarray}
where we suppressed the Lorentz indices
and fermionic terms.
We have further shown that this behavior is universal with respect to the choice of $G/H$.

We can also estimate the infrared contributions 
to the correlators. For the operator of dimension  $\Delta =4$, 
including the graviton mode, we obtain the nonanalytic behavior as
\begin{eqnarray}
\langle Str {\cal Y}_k [A,A][A,A] Str {\cal Y}_k^\dagger 
[A^\dagger,A^\dagger][A^\dagger,A^\dagger]
 \rangle_{IR} \sim k^4 \log k \ .
\end{eqnarray}
This is identical to those of conformal field theory.
Other gauge invariant operators with $\Delta =4$ are, for example,  
\begin{eqnarray}
tr {\cal Y}_k Z^4 \ ,
\label{KK}
\end{eqnarray}
 where $Z$ is a complex field such as $Z=A_a+iA_{a+1}$.
 $a$ runs over the directions transverse to $G$.
On $CP(2)$, $a=8,9$ and on $S^2 \times S^2$, $a=6,7,8,9$.
These operators correspond to the Kaluza-Klein modes and
are part of the supergravity modes.
Although they are no longer BPS operators off-shell, their
renormalization effects are finite as long as they carry finite momenta 
\cite{KTT}.
These operators (\ref{KK}) do not exhibit a nonanalytic behavior
which is peculiar to noncommutative gauge theory, since the trace in these operators is
not the symmetric ordered one but the ordinary one\footnote{
The symmetric trace is found  to be essential to obtain the nonanalytic ultraviolet
contributions in the two point functions\cite{KN}.}.
Other operators with $\Delta =4$, such as 
\begin{eqnarray}
tr {\cal Y}_k [A,A]Z^2 \ ,
\end{eqnarray}
also have no nonanalytic ultraviolet contributions specific to
noncommutative gauge theory.
We need more than two sets of commutators of $A$ in order to have 
the nontrivial effect from the symmetric trace.
Thus, we conclude that, in the gauge invariant $\Delta=4$ operators,
only the graviton type operators have the nonanalytic  
behavior of $1/k^2$.

We have also estimated the infrared contributions for the
operators which have higher dimensions $\Delta >4$.
The Kaluza-Klein modes receive the following infrared contribution
\begin{eqnarray}
\langle tr {\cal Y}_k Z^{\Delta} tr {\cal Y}_k^\dagger
 (Z^{\Delta})^\dagger \rangle_{IR} \sim k^{2 \Delta-4} \log k
\ . \label{WLC}
\end{eqnarray}
An interesting behavior is seen in the operators which 
do not belong to the supergravity multiplet.
For example,
the operators with dimension $\Delta$ receive 
the infrared contribution such as
\begin{eqnarray}
\langle Str {\cal Y}_k ([A,A])^{\Delta/2} Str {\cal Y}_k^\dagger
 ([A^\dagger,A^\dagger])^{\Delta/2}
 \rangle_{IR} \sim k^{2 \Delta-4} \log k \ ,
\end{eqnarray}
while they seem to receive the ultraviolet contribution as follows
\begin{eqnarray}
\langle Str {\cal Y}_k ([A,A])^{\Delta/2} Str {\cal Y}_k^\dagger
 ([A^\dagger,A^\dagger])^{\Delta/2}
 \rangle_{UV} \sim \frac{1}{k^{\Delta -2}}\ .
\end{eqnarray}
This behavior
might lead to some confinement mechanisms for these modes.

\section{Dual description in supergravity}
\setcounter{equation}{0}

In this section, we investigate the graviton propagators and
Kaluza-Klein modes in a dual supergravity. 
The supergravity solution which is dual to the $d=4$ euclidean 
noncommutative Yang-Mills theory
is associated with nonvanishing Neveu-Schwarz-Neveu-Schwarz $B$ fields 
\cite{HI,MR}.
In the low energy limit, 
the metric is written as
\begin{eqnarray}
\frac{1}{\alpha'} ds^2=R^2
(\frac{r^2}{1+a^4 r^4} d \vec{x}^2
+\frac{1}{r^2}d r^2+ d \Omega_5^2) \ ,
\end{eqnarray}
where $\vec{x}=(x_0,x_1,x_2,x_3)$ are the coordinates in the 
4 dimensions parallel to the brane.
The parameter $R$ is related with the rank of the gauge group $N$ and string coupling $g$ as
$R^4=4 \pi N g$. $a$ depends on the NS-NS $B$
field as
\begin{eqnarray}
B_{01}=B_{23}=\alpha' R^2 \frac{a^2 r^4}{1+a^4 r^4} \ .
\end{eqnarray}
By putting,
\begin{eqnarray}
a=\frac{\alpha'}{R} \ ,
\end{eqnarray}
and redefining the coordinate as 
\begin{eqnarray}
R^2 d \vec{x} \to d \vec{x} \ ,
\end{eqnarray}
we obtain
\begin{eqnarray}
ds^2=(\frac{R}{r})^2
(\frac{ d \vec{x}^2}{1+(\frac{R}{r})^4}
+dr^2+r^2 d \Omega_5^2
) \ , \label{met}
\end{eqnarray}
where we put $\alpha' =1$.
Other fields are expressed as
\begin{eqnarray}
&&e^{ \phi}= \frac{g(\frac{R}{r})^4}{1+(\frac{R}{r})^4} \ , \n 
&&B_{01}=B_{23}= \frac{1}{1+(\frac{R}{r})^4} \ ,  \n 
&&C_{01}=C_{23}= \frac{i}{g}  \frac{1}{1+
(\frac{R}{r})^4} \ ,  \n 
&&C = \frac{i}{g} (\frac{r}{R})^4 \ , \n 
&&F_{0123r}=\frac{4i R^{4}}{g} \frac{1}{(1+(\frac{R}{ r})^4)^2} 
\frac{1}{r^5} \ , \label{bg}
\end{eqnarray}
where $C$ denotes the R-R fields and $F$ denotes RR 5-form 
field strength.

The ordinary $AdS$ space is obtained by taking 
the commutative limit $R \to \infty$.
Then, the parameter $R$ becomes the radius of the $AdS$ space.
Since this background is considered to describe the strong coupling
region of noncommutative Yang-Mills, it is very useful to investigate various physics 
with noncommutative properties.
For example, in \cite{Gubser}, Gregory-Laflamme instability
\cite{GL1,GL2} for D0-D2
bound state in the black hole formation process
is discussed by investigating the thermodynamics in 
the decoupling limit of near-extremal D0-D2 system.
The relation between this instability and noncommutativity (matrix
model) is discussed in \cite{MS} where a fuzzy horizon emerges through
a gravitational collapse, which is a strong coupling phenomenon.

Anyway, to investigate the noncommutativity in string theory more
deeply,
it is essential to clarify the exact correspondence between the bulk theory and
boundary theory.
The guiding principle to connect these theories could be inferred from
the ordinary $AdS$/CFT correspondence. There 
the correlators in the boundary theory (CFT) are essentially
given by the boundary to boundary propagators in the bulk theory
\cite{GKP,Witten}.
With respect to the IR contributions, they should be smoothly connected with each other.
Other guiding principle, which we adopt in this paper,
is to reproduce the behavior of the UV origin in noncommutative gauge
theory,
that is, the behavior (\ref{gravuv}) which is obtained in \cite{KN}.
This behavior need not to be smoothly connected to that of commutative gauge theory 
since this is a specific effect in noncommutative gauge theory.

Our strategy is as follows: We aim to find the prescription which

i) is smoothly connected with that in the ordinary $AdS$/CFT correspondence
with respect to the infrared contributions and

ii) reproduces the ultraviolet contributions which is specific to noncommutative 
gauge theory due to UV/IR mixing effect.

The ambiguity of the prescription is where and how we should impose the boundary 
condition for the Green functions, which is closely related with the question 
where the brane exists in this background.
The coordinate $r$ in $AdS$ space corresponds to a length scale in the dual gauge
theory. Since we consider the noncommutative gauge theory,
the minimum length scale is the noncommutative scale.
Thus, we do not need to impose the boundary condition at $r= \infty$,
rather, a
natural idea is that we impose the boundary condition at the
characteristic scale in the background (\ref{met}), that is, $r=R$.
In order to observe why $r=R$ is the characteristic scale in this background,
we introduce the coordinate system which is conformally flat 
in the five dimensional subspace ($\vec{x},\rho$),
\begin{eqnarray}
ds^2&=&A(\rho) (d\vec{x}^2+d\rho^2)+R^2 d \Omega_5^2 \ , \n 
\rho&=&\int_R^r dr \sqrt{1+\frac{R^4}{r^4}} \ ,
\end{eqnarray}
where
\begin{eqnarray}
A(\rho) \sim \frac{R^2}{\rho^2} \ , \quad \rho \to \pm \infty \ .
\end{eqnarray}
We find that $A(\rho)$ has the maximum at $\rho=0$ ($r=R$) \cite{IIKK2,KTT}.

\subsection{Solutions}

Let us recall \cite{GH}, which is relevant to the solutions of a scalar mode in
this background.
Equation of motion for a scalar (and graviton) field
$\varphi$ becomes under the background (\ref{met}) and (\ref{bg}) as
\begin{eqnarray}
\frac{1}{\sqrt{g}} \partial_\mu \sqrt{g}g^{\mu \nu} e^{-2 \phi}\partial_\nu \varphi =0 \ .
\label{scleqn}
\end{eqnarray}
It is explicitly written as
\begin{eqnarray}
\left( \frac{5}{r} \pa_r+\pa_r^2 -R^4 k^2 (\frac{1}{r^4}+\frac{1}{R^4})
+\frac{\hat{L}^2}{r^2} \right) 
 \varphi 
 (\vec{k},r)=0 \ ,
\end{eqnarray}
where
\begin{eqnarray}
k\equiv |\vec{k}|=\sqrt{k_0^2+k_1^2+k_2^2+k_3^2} \ ,
\end{eqnarray}
which denotes the momentum along the brane and
$\hat{L}^2$ is the Laplacian on $S^5$.
The eigenvalue of Laplacian $\hat{L}^2$ is $-l(l+4)$ where
$l=0,1,2,\cdots$.
 Then,
for the $l$-th partial wave on $S^5$, we obtain
\begin{eqnarray}
\left( \frac{5}{r} \pa_r+\pa_r^2 -R^4 k^2 (\frac{1}{r^4}+\frac{1}{R^4})
-\frac{l(l+4)}{r^2} \right) 
 \varphi^{(l)} 
 (\vec{k},r)=0 \ .
\end{eqnarray}
The S wave on $S^5$ with $l=0$ 
corresponds to massless gravitons.
By changing the valuables as
\begin{eqnarray}
&&r=R e^{-z}, \n 
&&\varphi^{(l)} (\vec{k},r)=e^{2z} \psi^{(l)} (\vec{k},z) \ ,
\end{eqnarray}
the differential equation becomes
\begin{eqnarray}
\left(
\pa_z^2 +2(kR)^2 \cosh 2z -(l+2)^2
\right)
\psi^{(l)} (\vec{k},z)=0 \ . \label{mathieu}
\end{eqnarray}
Thus, the Mathieu's modified differential equation is obtained.
In ordinary $AdS$/CFT, $l$ is related to
the dimension of the operator $\Delta$ in the dual gauge theory as
$\Delta =l+4$.

Two independent solutions of this equation (\ref{mathieu}) are known as
\begin{eqnarray}
\frac{1}{r^2} H^{(1)} (\nu,z) \ ,  \quad 
\frac{1}{r^2} H^{(2)} (\nu,-z) \ ,
\end{eqnarray}
where $H^{(i)}$ denote the Mathieu functions.
The Floquet exponent $\nu$ is written in terms
of the combination $\lambda \equiv \frac{kR}{2}$.
It is related with $l$ as $\nu=l+2$ in the small momentum limit.
The explicit expressions for $\nu$ with $l=0,1,2$ are written as
\begin{eqnarray}
\nu&=&2-\frac{i}{3} \sqrt{5} \lambda^4 +\frac{7i}{108\sqrt{5}}
 \lambda^8+\cdots \ , \n 
\nu&=&3-\frac{1}{6} \lambda^4 +\frac{133}{4320} \lambda^8+\cdots \ , 
\n 
\nu&=&4-\frac{1}{15} \lambda^4 -\frac{137}{27000} \lambda^8+\cdots \ .
\end{eqnarray}

\subsection{Graviton modes}

The S wave is described by
the corresponding Floquet exponent as
\begin{eqnarray} \label{floquet2}
\nu=2-\frac{i \sqrt{5}}{3}(\frac{k R}{2} )^4
 +\frac{7i}{108\sqrt{5}}(\frac{k R}{2})^8+\cdots \ .
\end{eqnarray}
This mode corresponds to massless gravitons.
According to ordinary $AdS$/CFT,
the Green function in the bulk plays an important role to investigate the 
correspondence between the bulk theory and 
boundary gauge theory.
The Green function which is not divergent anywhere
in the entire region $0 < r < \infty$
is known as \cite{DG}
\begin{eqnarray}
G (r,r',k)&=&\frac{\pi}{4i}\frac{C}{A}
\frac{1}{r^2} H^{(2)} (\nu,-z)
\frac{1}{r^{'2}} H^{(1)} (\nu,z') \ , \quad r>r' \ , \n 
G (r,r',k)&=&\frac{\pi}{4i}\frac{C}{A}
\frac{1}{r^{'2}} H^{(2)} (\nu,-z') 
\frac{1}{r^{2}} H^{(1)} (\nu,z) \ , \quad r'>r \ .
\end{eqnarray}
where the normalization values $A$ and $C$ are determined by the 
asymptotic behaviors $z \to - \infty$ and $z \to \infty$ of the solution
$\frac{1}{r^2} H^{(1)} (\nu,z)$ \cite{GH}.
For the S wave, they are
\begin{eqnarray}
C&=&e^{i \pi \nu } -e^{-i \pi \nu} \ , \n 
A&=&\chi-\frac{1}{\chi}=\frac{\phi(-\nu/2)}{\phi (\nu/2)}-\frac{\phi
 (\nu/2)}{\phi (-\nu/2)} \ .
\end{eqnarray}
A meromorphic function $\phi (z)$ 
is defined as 
\begin{eqnarray}
\phi(z)&=&\frac{\lambda^{2z}}{\Gamma (z+r+1) \Gamma (z-r+1)} v(z) \ , \n 
v(z)&=&\sum_{n=0}^\infty (-1)^n \lambda^{4n} A_z^{(n)} \ , \n 
A_z^{(0)}&=&1 \ , \n 
A_z^{(q)}&=&\sum_{p_1=0}^\infty \sum_{p_2=2}^\infty \cdots \sum_{p_q=2}^\infty
a_{z+p_1} a_{z+p_1+p_2} \cdots a_{z+p_1+\cdots +p_q} \ , \n 
a_z&=&\frac{1}{(z+r+1)(z+r+2)(z-r+1)(z-r+2)} \ , \label{mero}
\end{eqnarray}
where $r \equiv {(l+2)}/{2}$.
This Green function behaves as $k^4logk$ for small $k$
and does not exhibit the $1/k^2$ behavior which is found
in dual noncommutative gauge theory  due to UV/IR mixing effect.
Even if we estimate the boundary contribution, we cannot reproduce 
such a behavior. Thus, we propose a new prescription which is different
from the proposal by Maldacena and Russo \cite{MR}.

First of all, we consider the region
\begin{eqnarray}
0<r,r'<R \ .
\end{eqnarray}
In this region, the solution $\frac{1}{r^2}H^{(1)}(\nu,z)$ is regular.
Thus, we can generalize the form of Green function as 
\begin{eqnarray}
G (r,r',k)&=&\frac{\pi}{4i}\frac{C}{A}\left(
\frac{x}{r^2} H^{(1)} (\nu,z) +\frac{1}{r^2} H^{(2)} (\nu,-z)\right)
\frac{1}{r^{'2}} H^{(1)} (\nu,z') \ , \quad r>r' \ , \n 
G (r,r',k)&=&\frac{\pi}{4i}\frac{C}{A}
\left(
\frac{x}{r^{'2}} H^{(1)} (\nu,z') +\frac{1}{r^{'2}} H^{(2)} (\nu,-z')\right)
\frac{1}{r^{2}} H^{(1)} (\nu,z) \ , \quad r'>r \ ,
\end{eqnarray}
where
$x$ is a constant to be determined by imposing
a boundary condition.
Although the term $\frac{1}{r^2 r^{'2}} H^{(1)} (\nu,z) H^{(1)} (\nu,z')$
diverges as $z \to -\infty$ ($r \to \infty $),
we allow the existence of this term since we consider the Green function 
in the finite $r$ region.

Next, we impose the boundary condition at $r =R$ and $r' \to 0$ as
\begin{eqnarray} \label{Neumann}
\pa_r G(r,r',k)|_{r=R}&=&0 \ , \n
G(r,r',k)|_{r' \to 0}&=&0 \ . \label{Neumann2}
\end{eqnarray}
The equation (\ref{scleqn}) is consistent with either Neumann or Dirichlet 
boundary condition at $r=R$ if we assume that it comes from
the following action.
\beq
\int d^{10}x \sqrt{g}g^{\mu \nu} e^{-2 \phi}\partial_\mu \varphi \partial_\nu \varphi  .
\eeq
We need to adopt the Neumann boundary condition
since the Dirichlet boundary condition gives the vanishing
boundary to boundary propagator.
This boundary condition is smoothly connected with the ordinary 
$AdS$/CFT prescription \cite{GKP,Witten} in the commutative limit as
it will be shown later.

In what follows, we estimate the non-analytic behavior of the Green function 
which determines the long distance behavior of the propagators
under the Neumann boundary condition.
$H^{(i)}$ are related with Floquet solutions $J (\nu,z)$ as
\begin{eqnarray}
H^{(1)}(\nu,z)&=&\frac{J(-\nu,z)-e^{-i\pi \nu}J(\nu,z)}{i \sin \pi\nu}\ ,
 \n 
H^{(2)}(\nu,z)&=&\frac{J(-\nu,z)-e^{i\pi \nu}J(\nu,z)}{-i \sin \pi\nu} \ ,
\label{jtoh}
\end{eqnarray}
and $J (\nu,z)$ is expanded in terms of Bessel functions as 
\begin{eqnarray}
J(\nu,z) = \sum_{n=-\infty}^\infty \frac{\phi (n+\frac{1}{2} 
\nu)}{\phi (\nu/2)}
J_n (\sqrt{q} e^{-z}) J_{n+\nu} (\sqrt{q} e^z) \ .
\end{eqnarray}
The explicit form of $J(\nu,z)$ is calculated as 
\begin{eqnarray}
J (\nu , z+\frac{i \pi}{2})& \sim & -\left(
(\frac{R}{r})^2 +\frac{1}{\chi_0} (\frac{r}{R})^2
\right) \frac{1}{2} \left(\frac{kR}{2} \right)^2 \n 
&&- \left(
(\frac{R}{r})^4 +\frac{1}{\chi_0} (\frac{r}{R})^4 -1-i\sqrt{5}
\right) \frac{1}{6} \left(\frac{kR}{2} \right)^4 + {\cal O } ( k^5 ) \ , \n 
J (-\nu , z+\frac{i \pi}{2}) &=&\frac{\phi (-\nu/2)}{\phi (\nu/2)}
J(\nu,-z-\frac{i\pi}{2}) \n 
&\sim & -\left(
(\frac{R}{r})^2 +\chi_0 (\frac{r}{R})^2
\right) \frac{1}{2} \left(\frac{kR}{2} \right)^2 \n 
&&- \left(
(\frac{R}{r})^4 +\chi_0 (\frac{r}{R})^4 -1+i\sqrt{5}
\right) \frac{1}{6} \left(\frac{kR}{2} \right)^4 + {\cal O} (k^5) \ ,
\end{eqnarray}
where $\chi_0$ is the leading order of the expansion of $\chi$ 
with respect to $\lambda=\frac{kR}{2}$ as
\begin{eqnarray}
\chi_0 \sim -\frac{2+i \sqrt{5}}{3} +{\cal O} (\lambda) \ .
\end{eqnarray}

$H^{(1)}(\nu,z)$ and $H^{(2)}(\nu,z)$ 
can be expanded as follows using the relation (\ref{jtoh}) 
\begin{eqnarray}
H^{(1)}(\nu, z+\frac{i \pi}{2}) &=&
\frac{2}{C} \left( J(-\nu,z+\frac{i\pi}{2})-e^{-i\pi \nu}
J(\nu,z+\frac{i \pi}{2}) \right) \n 
&\sim &-\frac{A_0}{C_0} \left( \frac{kr}{2} \right)^2 \left(
1+\frac{1}{3} (\frac{kr}{2})^2-(\frac{kR^2}{2 r})^2
\right) +{\cal O} (k^5)\ , \n 
H^{(2)}(\nu, -z-\frac{i \pi}{2}) &=&
-\frac{2}{C} \left( J(-\nu,z+\frac{i \pi}{2})-e^{i \pi \nu} 
J(\nu,z+\frac{i \pi}{2}) \right) \n 
&\sim &\frac{A_0}{C_0}\left( \frac{kR^2}{2r} \right)^2 \left(
1+\frac{1}{3} (\frac{kR^2}{2r})^2-(\frac{kr}{2})^2
\right) +{\cal O} (k^5) \ ,
\end{eqnarray}
where $C_0$ and $A_0$ are the leading order in the expansion of $C$ and
$A$ with respect to the momentum $k$
\begin{eqnarray}
C =C_0 +{\cal O} (k^5)
&\sim &\frac{2 \pi \sqrt{5}}{3} \left( \frac{k R}{2} \right)^4 +
{\cal O}(k^5) \ , \n 
A_0& \sim &-\frac{2\sqrt{5}i}{3} +{\cal O} (k) \ .
\end{eqnarray}
Note that there are no terms containing $\log k$ up to this order. 
Such terms emerge in the next leading order. 
Thus,
the boundary condition (\ref{Neumann}) determines $x$ as
\begin{eqnarray}
x \sim \frac{-6}{k^2R^2} + {\cal O} (1) \ .
\end{eqnarray}

In this way we find that the Green function for the small momentum behaves as
\begin{eqnarray}
G(r,r',k) & =&\frac{\pi}{4i}\frac{C}{A}\left(
\frac{x}{r^2} H^{(1)} (\nu,z) +\frac{1}{r^2} H^{(2)} (\nu,-z)\right)
\frac{1}{r^{'2}} H^{(1)} (\nu,z')|_{r=R,r'=R}
\n 
&\sim &\frac{3}{2k^2R^6} +{\cal O}(1)\ .
\end{eqnarray}
We have obtained the nonanalytic behavior of $1/k^2$. 
Such a contribution comes from the term
\beq
\frac{1}{r^2 r^{'2}} H^{(1)} (r)H^{(1)} (r'),
\eeq
which is dominant in $r \sim R$ region.
Thus, we interpret that this term
represents the ultraviolet contribution, which has been found in 
the Wilson line correlators (\ref{gravuv}) in the dual noncommutative 
gauge theory.
Note that the leading term of $H^{(1)}(r)$ is $O(r^2)$ so
the corresponding classical solution is $O(1)$ to the leading order.
The leading contribution vanishes when we impose the Neumann boundary condition
since we take the derivative with respect to $r$.
It is the reason why we obtain the $1/k^2$ type propagator in the end.
This effect is specific to the graviton type propagator and
does not take place in other Kaluza-Klein modes.
The infrared contribution $k^4 \log k$, 
which is considered to come from small $r$ region, 
is the same with that of the ordinary $AdS$ space.
We will discuss the infrared behavior in a commutative limit
later.

\subsection{Kaluza-Klein modes}

We will analyze the Green functions 
for the higher partial wave modes.
First, let us consider the Floquet exponent $\nu$ as
\begin{eqnarray}
\nu=3-\frac{1}{6} \lambda^4+ {\cal O}(\lambda^8) \ ,
\end{eqnarray}
which corresponds to the first excited partial wave mode.
The Green function is written as 
\begin{eqnarray}
G (r,r',k)&=&\frac{\pi}{4i}\frac{C}{A}\left(
\frac{x}{r^2} H^{(1)} (\nu,z) +\frac{1}{r^2} H^{(2)} (\nu,-z)\right)
\frac{1}{r^{'2}} H^{(1)} (\nu,z') \ , \quad r>r' \ , \n 
G (r,r',k)&=&\frac{\pi}{4i}\frac{C}{A}
\left(
\frac{x}{r^{'2}} H^{(1)} (\nu,z') +\frac{1}{r^{'2}} H^{(2)} (\nu,-z')\right)
\frac{1}{r^{2}} H^{(1)} (\nu,z) \ , \quad r'>r \ .
\end{eqnarray}
We impose the Neumann boundary condition 
(\ref{Neumann2}) to this Green function.
$x$ is determined by the boundary condition as 
\begin{eqnarray}
x\sim -1 \ .
\end{eqnarray}
The leading nonanalytic contribution of Green function 
is obtained as
\begin{eqnarray}
G (r,r',k)|_{r=R,r'=R} &=&\frac{\pi}{4i}\frac{C}{A}\left(
-\frac{1}{r^2} H^{(1)} (\nu,z) +\frac{1}{r^2} H^{(2)} (\nu,-z)\right)
\frac{1}{r^{'2}} H^{(1)} (\nu,z')|_{r=R,r'=R} \n 
&\sim &\frac{ 1}{54 R^4}+\frac{7 k^6R^2}{1536 }  \log \frac{kR}{2} \ .
\end{eqnarray}
A detailed calculation is given in Appendix.
$k^6 \log k$ comes from the both terms in the propagator.
This result implies that the leading nonanalytic behavior is due to both
ultraviolet and infrared contributions.
In the dual noncommutative gauge theory, Wilson line correlators 
with dimension $\Delta =5$ operator behaves as $k^6 \log k$,
which is seen in (\ref{WLC}).

Similarly, the Floquet exponent for the second level partial wave is given by
\begin{eqnarray}
\nu=4-\frac{1}{15} \lambda^4+ {\cal O}(\lambda^8) \ .
\end{eqnarray}
By imposing the Neumann boundary condition, we obtain
\begin{eqnarray}
x \sim -1 \ .
\end{eqnarray}
The leading nonanalytic contribution of the Green function is obtained as
\begin{eqnarray}
G (r,r',k)|_{r=R,r'=R} &=&\frac{\pi}{4i}\frac{C}{A}\left(
\frac{x}{r^2} H^{(1)} (\nu,z) +\frac{1}{r^2} H^{(2)} (\nu,-z)\right)
\frac{1}{r^{'2}} H^{(1)} (\nu,z')|_{r=R,r'=R} \n
&\sim &-\frac{ 1}{8 R^4}+k^8 R^4 \log \frac{kR}{2} \times (constant)
\end{eqnarray}
A detailed calculation can also be found in Appendix.
This behavior comes from both ultraviolet and infrared
contributions.
The corresponding operator $\Delta =6$ in the dual gauge theory
behaves $k^8 \log k$, which is seen in (\ref{WLC}).

\subsection{Commutative limit}

We will consider a commutative limit for the Green function.
The background approaches to $AdS_5 \times S^5$ in the limit
 $R \to \infty$ in (\ref{met}).
As the position of the boundary $R$ goes to infinity, our 
prescription smoothly 
goes over to that for the ordinary $AdS$/CFT as shown below.

In the finite $r$ region,
the two independent solutions satisfy 
\begin{eqnarray}
\frac{1}{r^2} H^{(2)} (\nu,-z) \gg \frac{1}{r^2} H^{(1)} (\nu,z) \ .
\end{eqnarray}
For the propagator dual to a $\Delta=4$ operator, such as the graviton propagator, 
the coefficient $x$ has been determined by the Neumann boundary condition as
\begin{eqnarray}
x \sim -\frac{6}{k^2R^2} +{\cal O}(1) \ .
\end{eqnarray}
For this operator, in the momentum region, 
\begin{eqnarray}
1 \ll \frac{1}{k^2R^2} \ll (\frac{R}{r})^4 \ ,
\end{eqnarray}
we find 
\begin{eqnarray}
\frac{1}{r^2} H^{(2)} (\nu,-z) \gg \frac{x}{r^2} H^{(1)} (\nu,z) \ .
\end{eqnarray}
Thus, the dominant part of the Green function which is well-defined in
the bulk is written as
\begin{eqnarray}
G(r,r',k)&=&\frac{\pi}{4i}\frac{C}{A}\frac{1}{r^2 r^{'2}}H^{(2)} (\nu,-z)
 H^{(1)}(\nu,z') \ , \quad r>r' \ , \n
G(r,r',k)&=&\frac{\pi}{4i}\frac{C}{A}\frac{1}{r^2 r^{'2}}H^{(2)} (\nu,-z')
 H^{(1)}(\nu,z) \ , \quad r'>r \ .
\end{eqnarray}
One can confirm that the nonanalytic behavior of this Green function 
becomes $k^4 \log k$, which is consistent with that in 
conformal field theory.
For $\Delta >4$, we can also consider the commutative limit and
obtain the identical behavior with that of conformal field theory by a similar discussion.

\section{Conclusion}
\setcounter{equation}{0}

We have investigated the Green function of the graviton mode in the 
type IIB supergravity background which is dual to noncommutative gauge theory.
We have shown that the leading nonanalytic term which comes from 
the ultraviolet contribution behaves as $1/k^2$ by imposing the
Neumann boundary condition on the Green function.
This contribution comes from the UV region which may be identified with
a characteristic effect seen
in noncommutative gauge theory.
This behavior is also seen in the graviton type Wilson line correlators in
the dual noncommutative gauge theory.
Thus, we have confirmed the existence of massless graviton 
on the noncommutative space from the viewpoint of dual supergravity picture.
In our analysis, we have required the following two conditions to fix
our prescription; namely 
i) it is smoothly connected with that in the ordinary $AdS$/CFT correspondence
with respect to infrared contribution and
ii) reproduces the ultraviolet behavior which is seen in noncommutative 
gauge theory due to UV/IR mixing effect.
The condition i) has determined the normalization factor
$C/A$ and condition ii) has determined $x$, which is the relative
coefficient of two independent solutions. In this way, we have determined the complete Green function.
We have also investigated Green functions of the operators with dimension
$\Delta =5,6$ with the Neumann boundary condition, 
and confirmed that there is no contribution which is specific to
noncommutative gauge theory.
This is consistent with the dual noncommutative gauge theory 
since $\Delta =5,6$ operators which are included in the supergravity
modes exhibit $k^{2 \Delta -4} \log k$ dependence
from the infrared contribution, 
which is identical with commutative gauge theory.

The adoption of the Neumann boundary condition on the propagator at $r=R$ 
is forced on us since
the Dirichlet boundary condition leads to the vanishing boundary to boundary
propagators. 
The remarkable outcome of our prescription is the existence of 
4 dimensional gravity which is consistent with the Newton's law a la Randall-Sundrum.

\begin{center} \begin{large}
Acknowledgments
\end{large} \end{center}
This work is supported in part by the Grant-in-Aid for Scientific
Research from the Ministry of Education, Science and Culture of Japan.
The work of S.N. is supported in part by the Research Fellowship of
the Japan Society for the Promotion of Science for Young Scientists.

\appendix

\section{Green function of $\Delta=5,6$ modes}
\setcounter{equation}{0}

The first order partial wave is described by 
the Floquet exponent $\nu$ as
\begin{eqnarray}
\nu=3-\frac{1}{6} \lambda^4+ {\cal O}(\lambda^8) \ .
\end{eqnarray}
For this mode, $C$ and $A$ are expanded by $\lambda =\frac{kR}{2}$ as
\begin{eqnarray}
C&=&e^{i \pi \nu } -e^{-i \pi \nu}  \n
  & \sim &\frac{\pi i}{3} \lambda^4 +{\cal O}(\lambda^{12}) \ ,
\n 
A&=&\chi-\frac{1}{\chi}=\frac{\phi(-\nu/2)}{\phi (\nu/2)}-\frac{\phi
 (\nu/2)}{\phi (-\nu/2)} \n
& \sim &-\frac{1}{\lambda^2}(1+\frac{1}{3} \lambda^4 \log \lambda )
(1+{\cal O} (\lambda^2)) \ .
\end{eqnarray}
Since we focus on the leading nonanalytic contribution,
we collect not only the terms in the leading order of $k$ but also
those in the leading order  of $log k$, 
even if they are not in the leading order with respect to $k$.
Then, $J(\nu,z)$, $J(-\nu,z)$, $H^{(1)}(\nu,z)$ and $H^{(2)}(\nu,-z)$
are calculated as
\begin{eqnarray}
J(\nu , z+\frac{i \pi}{2})
&\sim &\frac{i}{12}  \frac{k r^3}{R^2} 
(1+ \frac{1}{6} \lambda^4 \log \lambda )
\left(1+\frac{1}{4}(\frac{kr}{2})^2
+4 (\frac{kR^2}{2 r})^2+\cdots \right) \ ,
\n
J(-\nu, z+\frac{i \pi}{2}) &
\sim & \frac{i}{6} (\frac{k R^2}{2 r})^3 (1+\frac{1}{6} \lambda^4 \log
 \lambda )(1+{\cal O}(k^2)) \ ,
\end{eqnarray}
and
\begin{eqnarray}
H^{(1)}(\nu, z+\frac{i \pi}{2}) &=&
\frac{2}{C} \left( J(-\nu,z+\frac{i\pi}{2})-e^{-i\pi \nu}
J(\nu,z+\frac{i \pi}{2}) \right) \n
&\sim &\frac{i}{6 C_0} \frac{k r^3}{R^2} 
(1+\frac{1}{6} \lambda^4 \log \lambda )
(1+\frac{1}{4}(\frac{kr}{2})^2+4(\frac{kR^2}{2r})^2+\cdots ) \n
&&+\frac{2i }{36 C_0} (\frac{kR^2}{2 r})^3 \lambda^4 \log \lambda
\ , \n
H^{(2)}(\nu, -z-\frac{i \pi}{2}) &=&
-\frac{2}{C} \left( J(-\nu,-z-\frac{i \pi}{2})-e^{i \pi \nu} 
J(\nu,-z-\frac{i \pi}{2}) \right) \n
&\sim &-\frac{i}{6 C_0} \frac{kR^4}{r^3} 
(1+\frac{1}{6} \lambda^4 \log \lambda) 
(1+\frac{1}{4}(\frac{kR^2}{2r})^2+4 (\frac{kr}{2})^2+\cdots ) \n
&&-\frac{2i}{36 C_0} (\frac{kr}{2})^3 \lambda^4 \log \lambda
\ .
\end{eqnarray}
We determine $x\sim -1$ from the boundary condition (\ref{Neumann}).
In this way, the Green function is obtained as
\begin{eqnarray}
G (r,r',k)|_{r=R,r'=R} &=&\frac{\pi}{4i}\frac{C}{A}\left(
\frac{x}{r^2} H^{(1)} (\nu,z) +\frac{1}{r^2} H^{(2)} (\nu,-z)\right)
\frac{1}{r^{'2}} H^{(1)} (\nu,z')|_{r=R,r'=R} \n
&\sim &\frac{ 1}{54 R^4}+\frac{7 k^6R^2}{1536 }  \log \frac{kR}{2} \ .
\end{eqnarray}

The second order partial wave is described by the Floquet exponent as
\begin{eqnarray}
\nu=4-\frac{1}{15} \lambda^4+ {\cal O}(\lambda^8) \ .
\end{eqnarray}
$J(\nu,z)$, $J(-\nu,z)$, $H^{(1)}(\nu,z)$ and $H^{(2)}(\nu,-z)$
are calculated as
\begin{eqnarray}
J(\nu , z+\frac{i \pi}{2})
& \sim & -\frac{1}{5}  (\frac{r}{R})^4 (1+\frac{1}{15} \lambda^4 \log \lambda
 ) (1+{\cal O} (k^4)) \ ,
\n
J(-\nu, z+\frac{i \pi}{2}) &
\sim & \frac{1}{24} (\frac{k R^2}{2 r})^4 
(1+\frac{1}{15} \lambda^4 \log \lambda ) (1+{\cal O}(k^2))\ ,
\end{eqnarray}
and
\begin{eqnarray}
H^{(1)}(\nu, z+\frac{i \pi}{2}) &=&
\frac{2}{C} \left(J(-\nu,z+\frac{i\pi}{2})-e^{-i\pi \nu}
J(\nu,z+\frac{i \pi}{2}) \right) 
\n
&\sim &\frac{2}{5 C} (\frac{r}{R} )^4
(1+\frac{1}{15} \lambda^4 \log \lambda) (1-\frac{2}{15} \lambda^2 
+{\cal O}(k^4) )
\ , \n
H^{(2)}(\nu, -z-\frac{i \pi}{2}) &=&
-\frac{2}{C} \left(J(-\nu,-z-\frac{i \pi}{2})-e^{i \pi \nu} 
J(\nu,-z-\frac{i \pi}{2}) \right) \n
&\sim &-\frac{2}{5 C} (\frac{R}{r})^4 
(1+\frac{1}{15} \lambda^4 \log \lambda)(1-\frac{2}{15} \lambda^2
+{\cal O}(k^4))
\ . \label{mathi}
\end{eqnarray}
The normalization factors $A$ and $C$ are read off from the asymptotic 
behavior of the Mathieu functions. In $ \nu =4$ case, we need to
determine the next leading order of the expansion of $\lambda$,
which cancels the contribution of next leading order
of $\lambda$ in (\ref{mathi}).
\begin{eqnarray}
A&\sim &(\chi-\frac{1}{\chi})(1-\frac{4}{15} \lambda^2)
=\left(\frac{\phi(-\nu/2)}{\phi (\nu/2)}-\frac{\phi
 (\nu/2)}{\phi (-\nu/2)}
\right)(1-\frac{4}{15} \lambda^2) \n
& \sim &-\frac{24}{5\lambda^4}(1+\frac{2}{15} \lambda^4 \log \lambda)
(1-\frac{4}{15} \lambda^2) \n
C&=&(e^{i \pi \nu}-e^{-i \pi\nu}) \n
&\sim &-\frac{2 \pi i}{15} \lambda^4
\end{eqnarray}
We determine $x \sim -1$ from the boundary condition (\ref{Neumann}).
Thus, the Green function is obtained as
\begin{eqnarray}
G (r,r',k)|_{r=R,r'=R} &=&\frac{\pi}{4i}\frac{C}{A}\left(
\frac{x}{r^2} H^{(1)} (\nu,z) +\frac{1}{r^2} H^{(2)} (\nu,-z)\right)
\frac{1}{r^{'2}} H^{(1)} (\nu,z')|_{r=R,r'=R} \n
&\sim &-\frac{ 1}{8 R^4}+k^8 R^4 \log \frac{kR}{2} \times (constant) \ .
\end{eqnarray}

\end{document}